\documentclass[smallextended]{svjour3}
\usepackage{graphicx}
\usepackage{amsmath,amssymb}
\usepackage[numbers,square,compress]{natbib}
\newcommand{\ket}[1]{\left| #1 \right\rangle}
\newcommand{\bra}[1]{\left\langle #1 \right|}
\newcommand{\braket}[2]{\langle #1|#2 \rangle}
\journalname{Foundations of Physics}
\begin{document}
\title{How Quantum is Quantum Counterfactual Communication?}
\author{Jonte R. Hance \and James Ladyman \and John Rarity}
\institute{Jonte R. Hance \and John Rarity \at Quantum Engineering Technology Laboratories, Department of Electrical and Electronic Engineering, University of Bristol, Woodland Road, Bristol, BS8 1US, UK \\\email{jonte.hance@bristol.ac.uk}
\and James Ladyman \at Department of Philosophy, University of Bristol, Cotham House, Bristol, BS6 6JL, UK}
\date{Received: 03 Jun 2020 / Accepted: 02 Dec 2020}
\maketitle
\begin{abstract}
Quantum Counterfactual Communication is the recently-proposed idea of using quantum physics to send messages between two parties, without any matter/energy transfer associated with the bits sent. While this has excited massive interest, both for potential `unhackable' communication, and insight into the foundations of quantum mechanics, it has been asked whether this process is essentially quantum, or could be performed classically. We examine counterfactual communication, both classical and quantum, and show that the protocols proposed so far for sending signals that don't involve matter/energy transfer associated with the bits sent must be quantum, insofar as they require wave-particle duality.
\end{abstract}

\section{Introduction}
Quantum Counterfactual Communication is the combination of counterfactual circumstances (where ``things... might have happened, although they did not in fact happen"  \cite{Penrose1994Shadows}) with quantum physics, to send information between two parties without any matter/energy transfer associated with the bits sent (although in any practical implementation of it there is matter/energy transfer in the protocol as a whole). Given its interesting foundational implications, and potential for `unhackable' communication, it has excited massive interest in recent years \cite{Aharonov2019Modification,Al2016SalihQuantum,Arvidsson2016Communication,Arvidsson2019Postselection,Arvidsson2017Evaluation,Brida2012NohExperimental,Calafell2019Trace,Cao2014SalihDirect,Cao2017SalihComm,Chen2015Counterfactual,Chen2015SalihTripartite,Chen2016SalihDots,Gisin2013Photons,Griffith2016Path,Griffiths2017Reply,Griffiths2018Reply,Guo1999CryptoIFM,Guo2014SalihGates,Guo2014SalihEntangle,Guo2015SalihQI,Guo2017SalihCloning,Guo2018SalihEntSwap,Hance2019SalihChip,Jiang2011NohPractical,Li2013NohFlaw,Li2014NohEC,Liu2012NohDemons,Liu2014NohEaves,Liu2018SalihImprovement,Noh2009CounterfactualCrypto,Ren2011ExperimentNoh,Salih2013Invisible,Salih2013Protocol,Salih2014Tripartite,Salih2014ReplyVaidmanComment,Salih2014QubitOrig,Salih2016Qubit,Salih2018CommentPath,Salih2018Paradox,Salih2018Laws,Salih2018Erasure,Shenoy2013Wave,Shenoy2013NohPolar,Shenoy2013NohSemi,Shenoy2014NohCert,Shenoy2015NohSalihCat,Shukla2014OrthogSalih,Song2017NohBit,Sun2010NohQKD,Vaidman2013Past,Vaidman2014SalihCommProtocol,Vaidman2016Counterfactual,Vaidman2019Analysis,Wang2016NohCertAnalysis,Yang2016NohTrojan,Yin2010NohSecurity,Yin2012NohWeak,Zaman2018SalihBell,Zhang2012NohCCAttack,Zhang2012NohProof,Zhang2013NohDatabase,Zhang2014SalihImproved,Zhang2017ProbabilCounterfactual,Zubairy2014SalihMethod}.

What, if anything, makes these protocols essentially quantum \cite{Gisin2013Photons}? To answer this, we need to determine the underlying structure of classical counterfactual communication for comparison (Section \ref{Sect:ClassicalCounterfactuality}), and give a sufficient condition for a protocol to be quantum (Section \ref{sect:QAsNoClass}).

Section \ref{Sect:ProtEval} then examines the quantum counterfactual communication protocols proposed so far, to assess their non-classicality. We show what separates them from classical counterfactual communication protocols and how they meet the condition for being quantum.

We identify two essential differences between classical and quantum counterfactual communication. The first is that only one bit-value (e.g. `0') can be sent in a classical protocol without matter or energy transfer associated with the bit being sent. The second is that the two-value quantum protocols require wave-particle duality to be able to send either bit value of each bit sent.

\section{Classical Counterfactuality}\label{Sect:ClassicalCounterfactuality}

Counterfactual communication long predates quantum mechanics. For instance, in the Sherlock Holmes story, \textit{Silver Blaze}, Holmes infers a racehorse was abducted by its own trainer, as the stable dog didn't bark. As Holmes puts it, ``the curious incident of the dog in the night-time" was that the dog did nothing \cite{Doyle894Silver}. A more recent fictional example is the Bat-Signal. If there were a major crime being committed the Bat-Signal would appear in the sky, and so the Bat-Signal's absence counterfactually communicates to Bruce Wayne that all is well. Whenever we receive information from a sign's absence we are being signalled to counterfactually (e.g. the signal that an engine's components are functioning as they should is that the warning light is off).

Obviously in each of these cases a single bit is transmitted, and the bit value is signalled without the transfer of matter or energy. However, only one bit-value can be communicated by an absence in this way. Had a stranger kidnapped the racehorse, the dog would have barked, and energy would have have been transferred through the communication channel; correspondingly of course the Bat-Signal and other warning lights involve the transmission of energy when they are on. A sign's absence can transmit one value of a bit, only if the sign always occurs for the other bit value \cite{Maudlin2002Quantum}. This is counterfactual communication based on counterfactual inference.\footnote{It is not required that the signalled event's non-occurrence directly causes the the sign's absence, as there could be other common causal factors. Also a channel may be more reliable for signalling one bit value than for another.} Counterfactual inference is not rare but ubiquitous in everyday life and in science. For example, if there was an ether then the Michelson-Morley experiment would not have a null result.

The structure of classical counterfactual communication as above is as follows. Were A to happen, B would happen. B did not happen. Therefore A did not happen. B not happening is a signal that A did not happen, only because B happening signals that A happened.

Formally,

\begin{equation}
    \begin{split}
        A\supset B;\;\neg B;\; \therefore\neg A
    \end{split}
\end{equation}

Any instance of this structure in which B doesn't happen can be thought of as counterfactual communication of A's not happening. However, typically, we want a one-to-one correspondence between the signalling event A and the inferred event B, so we always recognise the inferred event's absence. For this, we need the further condition that, were A not to happen, B would not happen ($\neg A \supset \neg B$).

\section{Quantum as Non-Classical}\label{sect:QAsNoClass}
Next, to evaluate the proposed protocols we need a sufficient condition for something being quantum. There are many differences between classical and quantum physics. For optics (which all protocols so far have used), classical physics is everything up to and including Maxwell's equations. These formulate light as the evolution of waves whose intensity can be split continuously \cite{Griffiths2005Electrodynamics}. In contrast to this, in the quantum optics needed for many situations, we must consider light as photons \cite{Einstein1905PhotoElec}, which are quanta of the electromagnetic field that are detected as discrete packets of absorbed energy. Despite this discrete particle-like behaviour, in propagation photons retain wave-like properties such as interference. Therefore, in the context of optics it is appropriate to take a protocol to be quantum if it requires using both wave- and particle-like features by combining interference with single photon detection. The latter nullifies the splitting of light intensity across different detectors, and forces it to end in a single location.

\section{Protocol Evaluation}\label{Sect:ProtEval}
\subsection{Salih et al's Protocol}
\begin{figure}
    \centering
    \includegraphics[width=\linewidth]{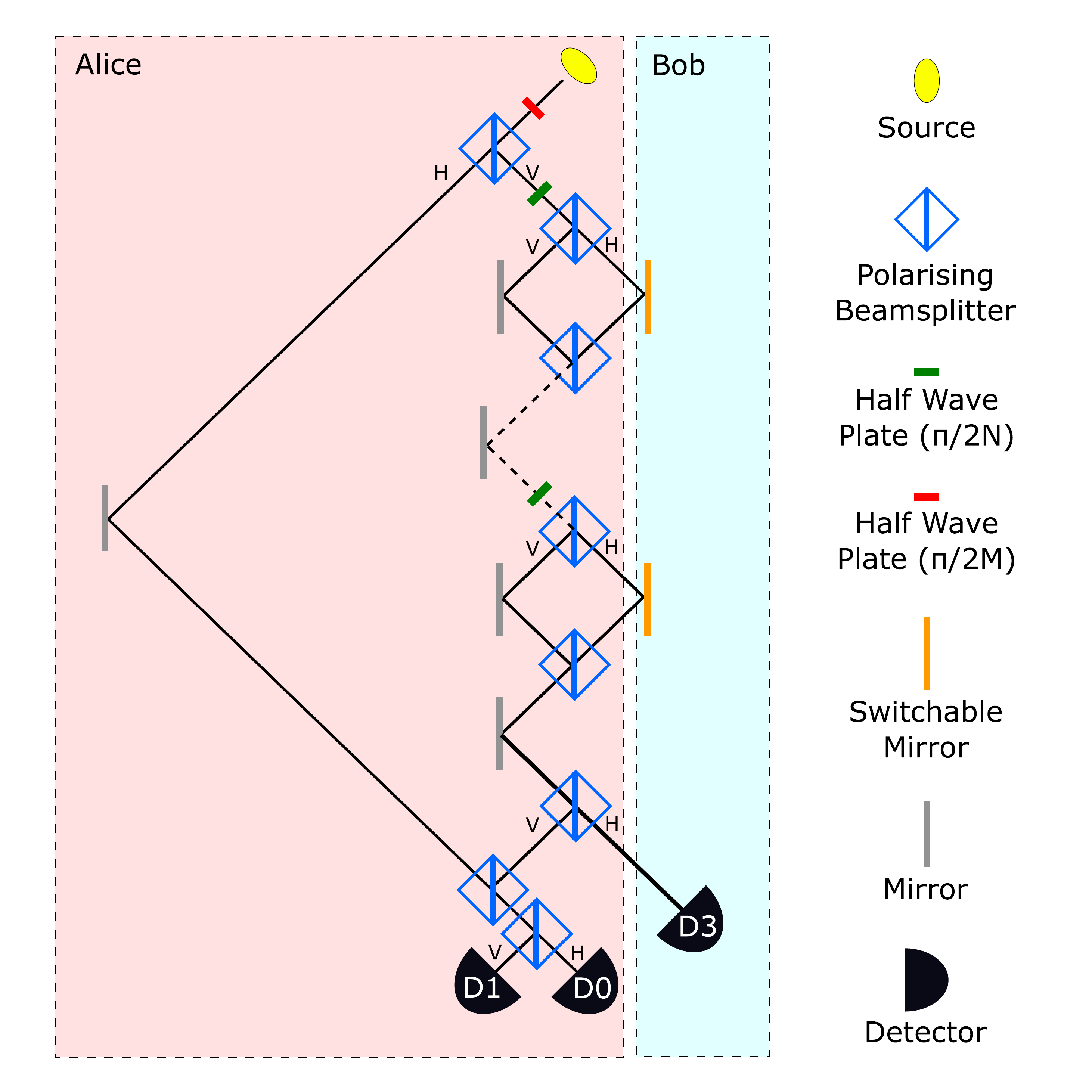}
    \caption{A schematic diagram of Salih et al's protocol for counterfactual communication, where, for every bit communicated, provably no photons have been to Bob. This version shows for one outer interferometer cycle ($M=1$), and multiple ($N$) inner interferometer cycles. The region of Alice is shown in pink, and the region of Bob in blue. Each Polarising Beamsplitter (PBS) reflects any vertically-polarised ($V$) light, and transmits any horizontally-polarised ($H$) light. The Half-Wave Plates (HWPs) rotate polarisation between horizontal and vertical polarisation unitarily, by an angle of either $\pi/2M$ (the HWPs before each outer interferometer) or $\pi/2N$ (the HWPs before each inner interferometer). This means, when Alice injects a horizontally-polarised photon from the source into her apparatus, the outer (left) path contains only horizontally-polarised light (labelled $H$), while a small amount of $V$-polarised light is created by the first HWP, and injected into the inner interferometer chain. In each inner interferometer, the left (Alice) path contains only $V$-polarised light, and the right (at Bob) only $H$-polarised light. If Bob does not block his paths, the chain of $N$ `$\pi/N$' rotations turn the $V$-polarised component in the inner interferometer chain to $H$, and so anything in the inner chain is sent to a loss channel $D_3$. This means Alice can only receive the photon if it went via her outer path, and so arrived at her $D_0$. If Bob blocks his paths, he absorbs this inner-chain $H$-polarised light, so the light is continually reset to $V$-polarised at the end of each inner interferometer, which stays on Alice's side, and reaches her $D_1$ as a $V$-polarised photon, having never travelled to Bob. A small number of photons are absorbed at Bob, reducing the efficiency. The probability of this happening decreases as $M$ and $N$ increase. Here, $M$ is 1, but in general $M\geq2$, $N\geq2$. The only way Alice's $D_1$ can click is if Bob blocks; and in the infinite limit of chained outer cycles, the only way her $D_0$ can click is if he doesn't. Unlike the classical case, both the `0' and `1' bit-values are received without energy transfer across the channel, and so both are sent counterfactually \cite{Salih2013Protocol,Salih2018Laws}.}
    \label{fig:Salih}
\end{figure}

Of the protocols proposed so far, only one has been shown counterfactual by both Weak Trace and Consistent Histories - Salih et al's \cite{Salih2013Protocol}\cite{Salih2018Laws}. We show this protocol in Fig.\ref{fig:Salih} and give a detailed description in the associated caption. Above we argued that a sufficient condition for a protocol to be quantum is that it requires both discreteness and path-interference - which, for light, only single photons can do. We now consider whether this protocol has to meet this condition in order to be counterfactual.

While the limit of many single photons may generate the same results as coherent states, the way in which they produce them differs. This is due to the discreteness discussed, which is not considered when using coherent states, but is when using Fock states (i.e. single photons).

In the quantum case, beamsplitters split a photon's probability amplitude between the two eigenstates that correspond to the photon going in each direction; in the classical case, they split the beam intensity (and field). As interference still occurs, when Bob does not block, waves on both sides still destructively interfere, so the light never returns to Alice. However, Bob's $D_3$ and Alice's $D_0$ both detect light simultaneously. Similarly, when he blocks, light goes to his blockers and Alice's $D_1$ simultaneously. Therefore, in both cases, as light goes between Alice and Bob, it is not counterfactual. While the amount of light going to Bob's $D_3$ may be infinitesimal for an infinite number of outer cycles, and that going to his blockers infinitesimal for infinite inner cycles, this is not the same as no light going there in either case - so, regardless the number of cycles, with classical light, the protocol isn't counterfactual. This may seem obvious, but many have not realised this and claimed this protocol could be performed classically (e.g. \cite{Gisin2013Photons}). 

The only way to avoid this is to force the light to end at only one point - to postselect, with information only travelling when nothing goes between Alice and Bob. Only single photons can do this. Therefore, the only way to make the protocol counterfactual is to use these, and so make the protocol quantum.

\subsection{Vaidman's Protocol} \label{Subsect:VaidmanProt}

\begin{figure}
    \centering
    \includegraphics[width=\linewidth]{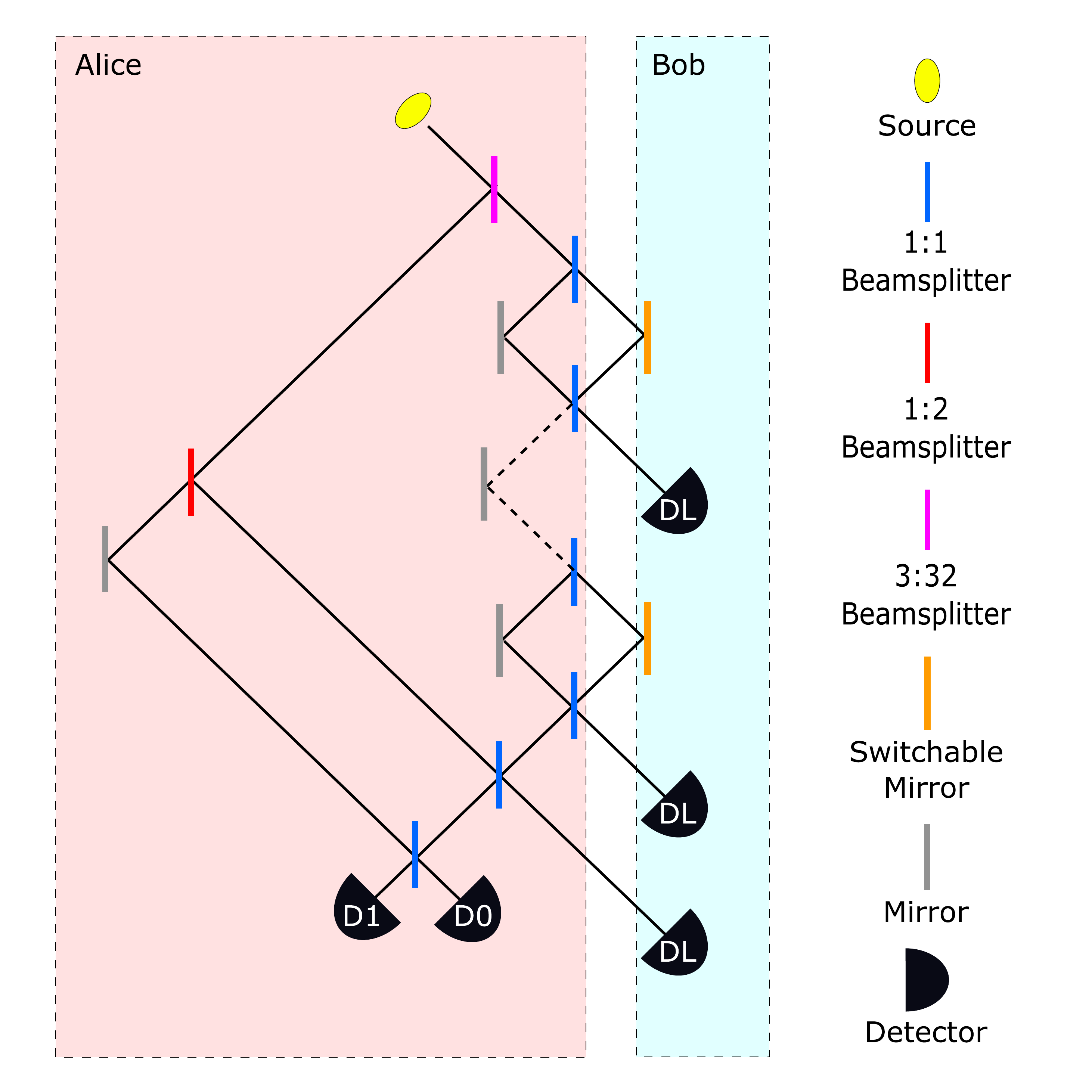}
    \caption{A schematic diagram of Vaidman's recent protocol for counterfactual communication \cite{Vaidman2019Analysis}. The region of Alice is shown in pink, and the region of Bob in blue. Unlike Salih et al's, it doesn't use polarised light, instead making use of ordinary (non-polarising) beamsplitters. By their design, when Bob doesn't block his side of the channel, each inner interferometer outputs into loss channels (here, the $D_L$s). This means (unless the photon is lost) the interferometer on the outer arm (starting with the red beamsplitter) always outputs at $D_0$. However, when Bob blocks, waves coming out of the inner interferometer chain negatively interfere at the final beamsplitter, causing the photon (if it stays at Alice) to go to $D_1$. The beam-splitting values given in this example make the losses equal, for two inner interferometers, regardless of whether or not Bob blocks his path (sends 1 or 0). We discuss this further in Appendix \ref{subsubsect:CFEqualLoss}.}
    \label{Fig:Vaidman2019}
\end{figure}

Alongside the proven protocol of Salih et al, Vaidman recently proposed one \cite{Vaidman2019Analysis} which, while not yet proven valid by Consistent Histories, has been shown to be valid by the Weak Trace criterion. We show this protocol in Fig.\ref{fig:Salih} and give a detailed description in the associated caption. Similarly to Salih et al's, it relies on both interference and single-photon detection - without the use of single photons, when Bob blocks the paths on his side of the inner interferometers, light could reach both Bob's blocker and Alice's detector $D_1$. Further, when Bob doesn't block, light could reach both the loss channels (marked `$D_L$' in Fig. \ref{Fig:Vaidman2019}) and Alice's $D_0$ - in both these cases, the protocol would definitely not be counterfactual.

Extrapolating from these protocols, any counterfactual protocol (valid or not) whereby the interference from Bob blocking or not blocking his side of an interferometer affects the destination of light on Alice's side, without them both simultaneously detecting that light, relies on both wave-like and particle-like properties. This means, all these protocols, from the Elitzur-Vaidman Bomb Detector, to Noh's counterfactual cryptography scheme, to even Arvidsson-Shukur et al's proposal (despite only sending one bit-value counterfactually), are essentially quantum by our definition above.

\section{Conclusion}\label{Sect:Conc}

We have shown Quantum Counterfactual Communication is essentially quantum. 
This confirms that both particle-like behaviour and path interference are necessary for schemes where \textit{both} bit-values are sent counterfactually. In all schemes demonstrated so far, this is the only way it is quantum (though protocols which send quantum information have been proposed theoretically \cite{Salih2014QubitOrig,Salih2016Qubit,Salih2018Paradox,Salih2020EFQubit,Salih2020DetTele}). Quantum Counterfactual Communication allows us to look at principles at the heart of the foundations of quantum physics - self-interference and counterfactual non-definiteness \cite{Hance2019Definiteness} - in a new and exciting way, and will hopefully motivate new thought experiments and experimental work based on this seemingly impossible phenomenon.

\begin{acknowledgements}
We thank Hatim Salih, Will McCutcheon, Paul Skrzypczyk and Robert Griffiths for useful discussions. This work was supported by the University of York's EPSRC DTP grant EP/R513386/1, and the Quantum Communications Hub funded by the EPSRC grant EP/M013472/1.
\end{acknowledgements}

\bibliographystyle{spphys}
\bibliography{ref.bib}

\appendix

\section{Path of a Quantum Particle}
To investigate Quantum Counterfactual Communication we need a criterion for a quantum particle being/not being somewhere. This will allow us to establish if a protocol is actually counterfactual.

\subsection{Classical Approach}
When considering where a quantum particle has been, our first instinct is to treat it as spatially local. However, as quantum phenomena are not solely particle-like, but as also have wave-like properties (as the Two-Slit Experiment shows \cite{Young1804Optics}), this is not necessarily the case. Therefore, we need a criterion for counterfactuality that takes into account quantum phenomena.

\subsection{No-Signal Approach}
Based on this need for a stronger criterion, a common view posed (typically on initially hearing of quantum counterfactual communication), is denying counterfactual communication is possible, by saying something can only be counterfactual if no information can pass between A and B. However, as we showed above, there are many classical protocols which are considered counterfactual communication, despite not fulfilling this criterion. Based on counterfactuality referring to something which ``might have happened, although they did not in fact happen", there is no reason to only call something counterfactual if no information can be transferred. Counterfactual communication is demonstrably possible - we just look at an extension of it.

\subsection{Density Matrix Approach}
Our first non-classical approach to a particle's location is using the entire quantum mechanical description of the system. We can do this by looking at the density matrix, which shows the whole of a state - it provides all the information that exists about a particle at a given moment. By associating them with physical positions, we can see the spread of the particle's possible locations. 

However, some parts of the density matrix correspond to paths lost when the wavefunction is collapsed at the protocol's end, and so information not being sent. We need some way to sort these possible paths into those where information is sent (that the particle could have been on when counterfactual communication occurred) and those where it is not. This requires post-selection (selection based on the final state they lead to), rather than just pre-selection from an initial state.

\subsection{Consistent Histories}

 To resolve this, we could use the Consistent Histories approach \cite{Griffiths1984Consistent}. 
This involves creating histories (tensored chains of projectors, each a way the system could evolve). A family is a set of these histories, that form a projective decomposition of the identity operator over the whole evolution time. By refining certain projectors (replacing them with projectors summing to them), we can model a situation. These histories are consistent if they all mutually commute - if they do not, it is meaningless to ask which history was the `correct' model for the system's evolution, as we cannot assign positive relative  probabilities
 \cite{Griffiths2019SEP}.

Therefore, we can say a particle has not gone between Alice and Bob when all histories where it travels between the two (where information is sent), have probability zero. This requires us to analyse all histories in this family, as Griffiths does for several protocols \cite{Griffith2016Path}. In essence, if a particle can go between Alice and Bob when information is transmitted, by Consistent Histories it is not counterfactual; 
if not all the histories in the family are consistent with one another, it is meaningless to talk about the counterfactuality of the situation.

\subsection{Weak Trace}

\begin{figure}
    \centering 
    \includegraphics[width=\linewidth]{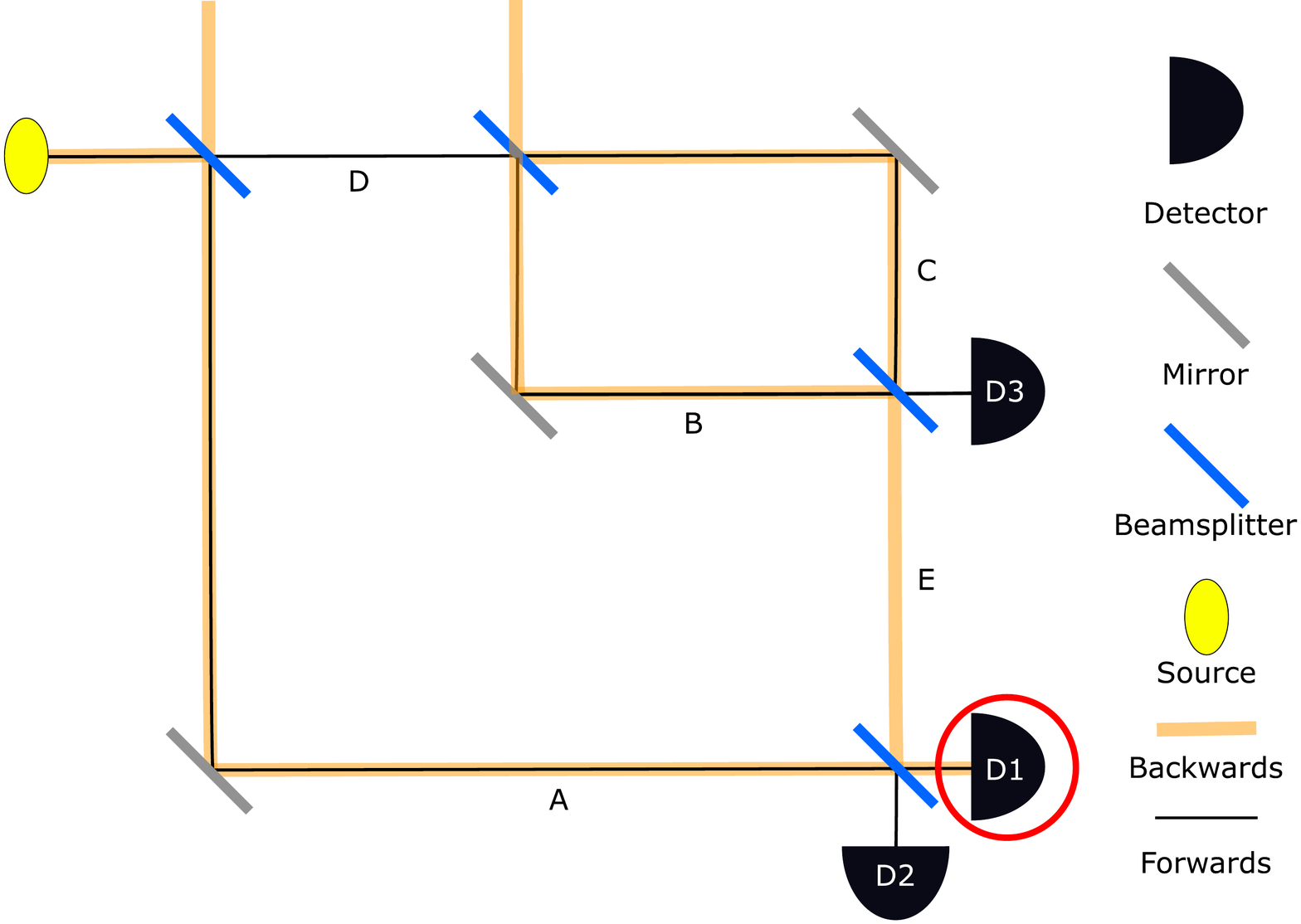}
    \caption{The Two-State Vector Formalism applied to a nested interferometer. Forwards-travelling paths are marked by thin black lines, and backwards-travelling paths by thick orange lines. Though no forward or backwards travelling path goes from the source, to Bob (along path C) and into D2, they do overlap over C, meaning there is a weak trace at Bob. This illustrates the peculiar property of the TSVF where particles can jump between regions (e.g. between the inner interferometer and the outer arm) \cite{Vaidman2013Past}.}
    \label{fig:TSVFDiag}
\end{figure}

Next, we consider weak measurement \cite{Aharonov1988Weak}, which examines the state between measurements, without collapsing it. Weak measurement involves lightly coupling a system to a measuring device, so while little information is gathered over one run of the system, over many runs a probability distribution is obtained. This contrasts with Von Neumann measurements, which cause a system to collapse into an eigenstate of the measured operator. Weak measurement allows us to collect information that would be lost were the system strongly measured \cite{Tamir2013Intro}. To do this we apply the expectation value of the evolution operator to the initial state, which we can interpret as evaluating all forward-evolving paths from that state.

However, rather than working forwards, can also work back from a given result (post-select), to investigate the paths the system may have evolved through. If we pre- and post-select like this, we say a particle leaves a weak trace (indicating possible presence) wherever this weak measurement value is non-zero. To approximate this trace, we can trace the initial vector forward, and the final vector backward, in time, and see where they overlap. If we represent the evolution of the state by operator $\hat{O}$, we get this approximate value as

\begin{equation}
    O_w=\frac{\bra{\psi_f}\hat{O}\ket{\psi_i}}{\braket{\psi_f}{\psi_i}}
\end{equation}

where $\ket{\psi_i}$ is the initial state of the particle, and $\ket{\psi_f}$ the final. This approximation of the weak value (to the first order in trace size, $\mathcal{O}(\epsilon)$) is the Two-State Vector Formalism (TSVF), as it goes from/to two states - that at the beginning of the protocol, and that at the end. This gives both pre- and post-selection needed.

If an operator returns a non-zero TSVF value, there is to $\mathcal{O}(\epsilon)$ a weak trace along the path it describes. If we trace the paths a quantum particle could evolve along from its initial, and those it could have come from to get to its final, state, there is a weak trace where they overlap - so we cannot say the particle was not there \cite{Vaidman2013Past}.

However, this has unintuitive results. While, with Consistent Histories, a path needs to link the initial and final states, here, it only requires paths from the initial and final states overlap at some point. This means Bob can have a weak trace on his side of the transmission channel, without any in the channel itself. This was demonstrated using weak measurements in nested Mach-Zehnder Interferometers (MZIs) by Danan et al \cite{Danan2013Asking}. If one accepts the weak trace as a valid indicator of a particle's path, this leads to peculiarities - such as particles jumping discontinuously between locations \cite{Aharonov2017Disappearing}. This caused Sokolovski to doubt the formalism, in favour of continuous paths. \cite{Sokolovski2013ReallyWeak,Englert2017Revisited,Sokolovski2018Simpler}. However, these peculiar results don't contradict standard quantum theory \cite{Peleg2018CommentRevisited}.

A more compelling counterargument is that the TSVF ignores the non-$\mathcal{O}(\epsilon)$ weak trace, and so does not give the particle's entire path. Vaidman admits this, saying the TSVF only gives the weak trace to $\mathcal{O}(\epsilon)$. Further, analysis of Danan et al's data shows smaller, $\mathcal{O}(\epsilon^2)$ peaks not visible in their original presentation \cite{Sokolovski2017Plain}. Vaidman explains this by saying the non-local trace on any particle is also of $\mathcal{O}(\epsilon^2)$, and so this applies in any set-up, even if objects are physically separated. Further, as there are no non-local interactions in nature, this non-local weak trace cannot be strong enough to mediate any effects, so neither can a local second-order weak trace \cite{Vaidman2013Past}. Despite this, Vaidman still claims this constitutes a weak trace for Salih et al's one-outer-cycle protocol \cite{Vaidman2019Analysis}.

\section{Quantum Counterfactual Communication Protocols}

In this Appendix, we look at Quantum Counterfactual Communication protocols proposed so far. Since Elitzur and Vaidman first discovered quantum counterfactuality \cite{Elitzur1993Bomb}, and Kwiat et al allowed loss to be made effectively nil \cite{Kwiat1995IFM}, researchers have tried to exploit it for communication. Despite this, all protocols until recently fell into three broad categories: where communication is counterfactual only for one bit-value; where photons travel between Alice and Bob, but in the opposite direction to the information passed between them; and where no photons pass between Alice and Bob when information flows, but the error/loss rates vary with the bit-value Bob sends.

\subsection{Elitzur-Vaidman Bomb Tester}

\begin{figure}
    \centering 
    \includegraphics[width=\linewidth]{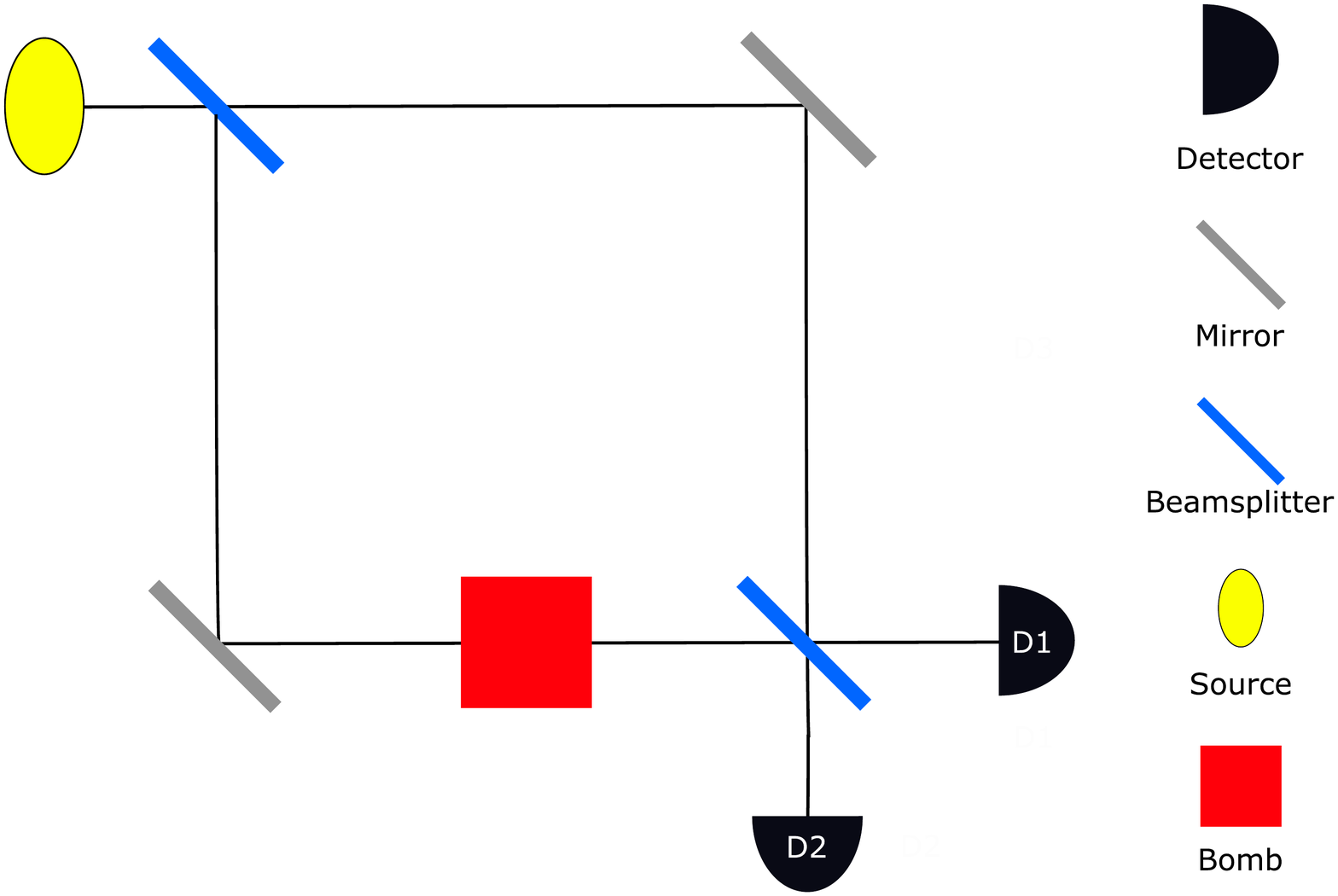}
    \caption{The Elitzur-Vaidman Bomb Tester. A photon is emitted from the source (top-left), enters the balanced Mach-Zehnder interferometer, and is spread across both paths equally. If the bomb is faulty, the photon recombines at the second beam-splitter, and always enters $D_1$. If the bomb works, and is activated, it destroys the set-up. If the bomb would work, but the photon went down the bomb-free path, the photon has a 50:50 chance of going to either detector.}
    \label{fig:EVDiagram}
\end{figure}

All quantum counterfactual communication protocols stem from the Elitzur-Vaidman Bomb-Detector \cite{Elitzur1993Bomb}. Here (Fig.\ref{fig:EVDiagram}), an MZI has a potentially faulty bomb along one of its paths, which can only be detonated by a non-demolition single-photon detection. 

If the photon goes along the bomb's side of the MZI (and the bomb works), it detonates, and the photon (and everything else) is destroyed; if the bomb is faulty, the photon travels to the merging beamsplitter normally. However, if the photon traverses the other side, the bomb working changes the interference pattern, making it able to go to a detector it previously couldn't access. This allows us to test if the bomb would have worked, without detonating it, by putting it on the path the photon could have, but did not, travel down. Unlike classical counterfactual communication, both options are transmitted counterfactually - the photon's path is from the source to the detectors without going via the object (bomb) under evaluation. However, it is not necessarily always counterfactual. This is as, while the photon \textit{can} carry the information without going via the bomb, it does not necessarily have to. This means the Bomb-Tester is not fully counterfactual.

\subsection{Counterfactual only for One Bit}

Here, the protocol is counterfactual for one bit-value, but the photon goes between Alice and Bob for the other.

\begin{figure}
    \centering
    \includegraphics[width=\linewidth]{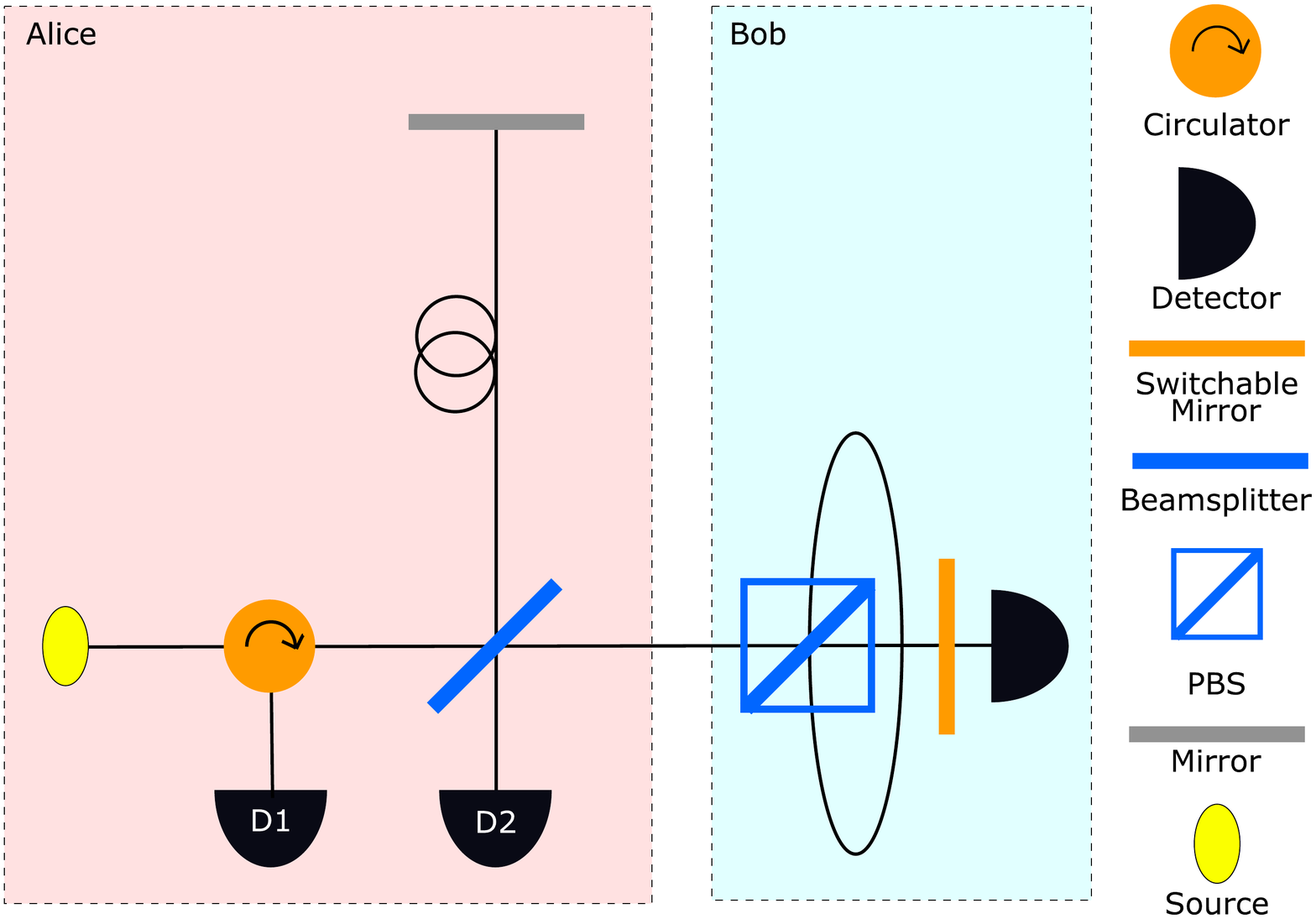}
    \caption{Noh's counterfactual cryptography protocol - Alice randomly polarises a photon, which passes through a beam-splitter, with one of the outputs going to Bob. Bob uses a PBS and delay to time-separate possible polarisations arriving it him, and picks one to reflect and one to absorb. If the photon is Bob's reflect-polarisation, it reflects back, and interferes into Alice's $D_2$. However, if the photon is his absorb-polarisation, it is sent into his detector. If this clicks, the protocol is aborted; if not, the photon goes into $D_1$ \cite{Noh2009CounterfactualCrypto}.}
    \label{fig:Noh2009}
\end{figure}

The first of this type of protocol, and indeed the first Quantum Counterfactual Communication protocol proposed was Noh's (Fig.\ref{fig:Noh2009}) \cite{Noh2009CounterfactualCrypto}, (barring Guo's E-V Bomb Detector adaption, where photons travel between Alice and Bob for both bit-values \cite{Guo1999CryptoIFM}). For matched polarisations, if Alice gets a click, the photon has remained on her side. However, for orthogonal polarisations, it has both been to, and returned from, Bob - so is not counterfactual. Despite this, the work generated a lot of interest \cite{Sun2010NohQKD,Yin2010NohSecurity,Jiang2011NohPractical,Ren2011ExperimentNoh,Brida2012NohExperimental,Liu2012NohDemons,Yin2012NohWeak,Zhang2012NohCCAttack,Zhang2012NohProof,Shenoy2013NohPolar,Shenoy2013NohSemi,Li2013NohFlaw,Zhang2013NohDatabase,Li2014NohEC,Liu2014NohEaves,Shenoy2014NohCert,Shenoy2015NohSalihCat,Wang2016NohCertAnalysis,Yang2016NohTrojan,Song2017NohBit,Shenoy2013Wave}. While plenty of these focus on reducing loss by reducing the proportion of the photon sent to Bob \cite{Sun2010NohQKD}, this must be non-zero for the protocol to function. Therefore, the system will never be fully counterfactual.

\subsection{Information and Photon Travel in Opposite Directions}

In the next category, the photon can cross the channel, but in the opposite direction to the information being sent. Here, for one bit-position the photon destructively interferes across the quantum channel, keeping it at Alice, and for the other, it constructively interferes, allowing it through to Bob. Based on if she detects a photon, Alice can determine which bit Bob sent.

The only protocol of this sort is Arvidsson-Shukur et al's. They propose a device formed of chained MZIs, which use the Quantum Zeno effect to (for many MZIs) keep the photon at Alice if Bob blocks, and force it to go to Bob if he does not \cite{Arvidsson2016Communication} - identically to Kwiat et al's Interaction Free Measurement protocol \cite{Kwiat1995IFM}. Ignoring the high chance of Alice wrongly believing Bob did not block (due to the necessarily finite number of MZIs causing some chance of Bob's blocker absorbing the photon), the photon still travels at the same time as the information. Waves carrying information in the opposite direction to travel is a well-known classical phenomenon \cite{Vaidman2019Analysis}. Therefore, this protocol only seems quantum when you consider light as local - where, for one possibility, the photon travels from Alice to Bob. This obviously creates a weak trace at Bob, and so is not counterfactual. Arvidsson-Shukur et al attempt to advocate their protocol by saying it tolerates error better than others \cite{Arvidsson2017Evaluation}, and by calling other protocols classical using a classical model with Alice and Bob having extra, non-trivial resources (e.g. a shared clock) \cite{Arvidsson2019Postselection}. However, they give no reason to view their protocol as true Quantum Counterfactual Communication, and so their work with Calafell et al \cite{Calafell2019Trace} just demonstrates classical counterfactuality.

\subsection{Photon only Travels Erroneously}
The next set of protocols have Alice receiving a photon for both bit values, which has never been to Bob. This means the photon cannot go to Bob when Alice gains information, as then Alice would be unable to see what was sent. Therefore, when photons go to Bob, the protocol is aborted and retried, creating a source of loss. 

\subsubsection{Unequal Losses}
For these protocols, this loss varies with the bit-value sent. This leads us to ask if Alice can, by knowing loss probabilities, guess the bit Bob sends solely based on if any of her detectors light up. This would make the same as the last category. We can also ask if this post-selection is to blame for any peculiar effects observed, even with this loss enforced according to the protocol.

Salih et al's 2013 protocol was the first claiming to be fully counterfactual for both bit values \cite{Salih2013Invisible,Salih2013Protocol}. It is formed of a chain of outer interferometers, each containing a chain of inner interferometers (see Fig.\ref{fig:Salih}). However, Vaidman claimed this was not counterfactual, when assessed by the Weak Trace criterion (see Fig.\ref{fig:TSVFDiag}) \cite{Vaidman2013Past}. The TSVF gives a weak trace on Bob's side of the channel when Bob does not block this, meaning we cannot say the photon was not there \cite{Vaidman2014SalihCommProtocol,Vaidman2016Counterfactual}. However, this is only for the simplest (polarisation-free) form of the protocol. Using polarisation (as shown in Fig.\ref{fig:Salih} and \cite{Salih2018Laws}) avoids a weak trace on Bob's side by ensuring the only waves that go to Bob are H-polarised, which are lost via $D_3$ on Bob's side, restarting the protocol \cite{Salih2014ReplyVaidmanComment}. Then, when Bob does not block, the differences in polarisation between the forward- and backward-travelling states keep them separate on Bob's side - giving no Weak Trace there. This was tested practically through weak measurement, using Danan et al's method \cite{Danan2013Asking} and shown to have no weak trace from Bob's side visible at Alice's detectors. Griffiths also claimed it is not counterfactual by Consistent Histories, as a history with a non-zero probability could be traced to Bob's side and back when Bob does not block \cite{Griffith2016Path,Griffiths1984Consistent,Griffiths2017Reply}. However, again, Griffiths only considered physical paths, rather than polarisations, which provide an extra degree of freedom \cite{Salih2018CommentPath}. Griffith later showed, when using more than one outer cycle, the family became inconsistent, and so calling the protocol counterfactual was meaningless \cite{Griffiths2018Reply}. However, Salih notes the final outer cycle is counterfactual, while identical earlier ones are meaningless, which seems paradoxical \cite{Salih2018Paradox}. Once Salih et al published their protocol, various implementations began to appear \cite{Cao2014SalihDirect,Guo2014SalihEntangle,Guo2014SalihGates,Salih2014Tripartite,Zhang2014SalihImproved,Shukla2014OrthogSalih,Zubairy2014SalihMethod,Chen2015Counterfactual,Chen2015SalihTripartite,Guo2015SalihQI,Shenoy2015NohSalihCat,Al2016SalihQuantum,Chen2016SalihDots,Salih2014QubitOrig,Salih2016Qubit,Cao2017SalihComm,Guo2017SalihCloning,Guo2018SalihEntSwap,Salih2018Erasure,Liu2018SalihImprovement,Zaman2018SalihBell,Salih2018Paradox,Hance2019SalihChip,Hance2020CFGhostImg,Salih2020EFQubit,Salih2020DetTele}. While many of these don't make use of polarisation, some do.

Zhang et al proposed a protocol, based on Salih et al's, for probabilistic counterfactual communication. Their protocol isn't always counterfactual, but they claim the chance of the photon being at Bob can be reduced to nil, and losses (from noise/blocking) are lower \cite{Zhang2017ProbabilCounterfactual}. However, they assume photons only trace one path, which isn't always true - so it is not counterfactual.

Despite originally claiming counterfactual communication of both bit-values was impossible, at roughly the same time as Salih et al defended their protocol using polarisation, Vaidman, alongside Aharonov, released a protocol allowing just this \cite{Aharonov2019Modification}. This method is effectively the same as in Salih et al's original protocol - however, to avoid a weak trace in this set-up, where there is no polarisation degree of separation, at least two inner interferometers are needed. Alongside this, Aharonov and Vaidman make repeated reference to a double-sided mirror in the protocol; but all this does is connect the two inner interferometers, and fold the outer path to reduce physical space used, and so it is irrelevant to the protocol's counterfactuality. Also, unlike Salih et al's, this protocol is not counterfactual by Consistent Histories \cite{Salih2018Paradox}.

\subsubsection{Equal Losses}\label{subsubsect:CFEqualLoss}
Shortly after publishing with Aharonov, Vaidman created another weak trace-free counterfactual communication protocol. For one outer cycle, this protocol avoids the risk of an erroneous reading that Salih et al's, and their earlier, protocol has \cite{Vaidman2019Analysis}. It is again based on a chained MZI set-up, but uses interference from light passing through the inner interferometer chain to alter which detector the photon ends up at when Bob blocks (see Section \ref{Subsect:VaidmanProt}). This allows Alice, when she receives a bit, to be certain it is the same value Bob sent. Another benefit of the protocol is that, for certain beam-splitter values, losses were the same whether or not Bob blocked. This means Alice cannot infer if Bob blocked, just based on if she receives a photon. Therefore, the protocol cannot be reduced to the information and photon travelling simultaneously, in opposite directions, and so it seems counterfactual. However, it remains to be seen if it is fully counterfactual by the Consistent Histories criterion.

\end{document}